\documentclass[preprint,review,12pt]{elsarticle}
\usepackage{epsfig}
\usepackage{amssymb}
\usepackage{amsthm}

\begin{document}

\begin{frontmatter}

\title{Evolution of Skyrmion Crystals in Fe$_{0.5}$Co$_{0.5}$Si-Like Quasi-Two-Dimensional
 Ferromagnets   Driven by  External Magnetic  Field and Temperature}

\author{Zhaosen Liu$^{a,b}$\footnote{Email: liuzhsnj@yahoo.com},
Tiantian Huan$^{b}$, Hou Ian$^{b,c}$\footnote{Email:
houian@umac.mo}}

\date{\today}
\vspace{1cm}

\begin{abstract}
Magnetic skyrmions have attracted great research interest in
recent years due to their exotic physical properties, scientific merit and potential applications in modern technology. Here,   we apply   a quantum computational method to  investigate the spin configurations of Fe$_{0.5}$Co$_{0.5}$Si-like
quasi-two-dimensional ferromagnetic system   with co-existence  of
Dzyaloshinsky-Moriya and Heisenberg exchange interactions. We find
that within  a weak  magnetic field perpendicular to   the film plane, skyrmion crystal (SkX) of hexagonal-close-packed pattern can be induced, the  spin configurations  evolve  with applied magnetic  field   and temperature.  This quantum model, if scaled, is able to qualitatively reproduce the experimental results of SkX with long periodicity.  Especially, when the skyrmion size is
around a few nano-meters in diameter, \emph{or more general,  when the characteristic length  of the material approaches the lattice scale}, the quantum model is expected to be more accurate than the classical ones.
\end{abstract}


\end{frontmatter}

\section{Introduction}

 Skyrmions,  a novel type of topological magnetic structures
 observed  in solids,  have attracted intensive research interest
 in recent years
\cite{Bogdanov01,Lee09,Schulz12,Jonietz10}. The concept of
skyrmion was originally introduced   in nuclear physics to
describe a localized, particle-like  configuration in field theory
about half century ago \cite{01Skyrme}.  However,  these
topological structures were  lately  predicted to exist in
magnetic materials based on micromagnetic model
\cite{1Bogdanov,2Bogdanov}, and experimentally observed  in bulk
materials, such as   MnSi   with the B20 crystal structure
\cite{Pappas09,Muhlbauer09}, Fe$_{1-x}$Co$_x$Si and FeGe thin
films \cite{Yu10, Yu11}, multi-layer systems
\cite{8Crepieux,9Bode,16Heinze,17Romming,Dupe,Dupe2}, as well as
insulating magnet Cu$_2$OSeO$_3$ \cite{5Seki}. Magnetic skyrmion
crystal   (SkX)  textures   are mainly induced by  the Dzyaloshinsky-Moriya interaction (DMI)  \cite{6Moriya,7Dzyaloshinskii}. This chiral interaction  exists  in the systems with broken structural inversion symmetry, and is present  at the surfaces  of thin films or the  interfaces of multi-layers. Since transition-metal interfaces  are essential in spintronic devices, the discovery of SkX on these interfaces \cite{16Heinze,17Romming,Dupe} has aroused great research interest in using them as novel data-storage device concepts \cite{20Fert,21Sampaio}.

In three-dimensional (3D) materials,  SkX is usually stable  in  a
rather narrow $T$-region immediately below  the transition
temperature as observed in neutron scattering experiment conducted
on MnSi bulk magnet \cite{Muhlbauer09}.
 However in  two-dimensional (2D) systems,   SkX can exist over a
 wide  temperature region \cite{Yu11,Han10}. For instance, using Lorentz transmission electron microscopy, Yu et al.~observed that when Fe$_{0.5}$Co$_{0.5}$Si thin film   is  placed in  external magnetic fields,  the SkX appears and persists down to almost zero temperature \cite{Yu10}.  These skyrmions crystalize in  the  hexagonal-close-packed (HCP) pattern with a periodicity of about 90 nm.

 So far, various descriptions  for magnetic skyrmions have been proposed.
 Most of them  include the notion of topology  as defined in  micromagnetics  where the  continuous model is applied. There, a magnetic skyrmion can be described with  a non-zero  integer value of the topological index,  referred to as  topological charge, or   winding number. The topological charge of this spin texture can be expressed as
\begin{equation}
Q = \frac{1}{4\pi}\int\int
\vec{m}\cdot\left(\partial_x\vec{m}\times\partial_y\vec{m}\right)dxdy
\label{Q}\;.
\end{equation}
 where   $\vec{m}$  is a unit vector specifying the direction of the local magnetic moment, and the integral is taken over  the space occupied by the skyrmion.

 However, we must bear in mind that the concept of topology can only  be rigorously applied in continuous models to  infer the stable spin structures
  \cite{wikipedia}. At a size scale less than a few nanometers,  or more general, \emph{as the   spin-spin correlation length of the material  approaches the unit cell scale},  the spin textures of magnetic materials become noncontinuous  due to the discretization of the atomic lattice. For instances,   the skyrmions formed at    interfaces are often a few nanometers in diameter  \cite{16Heinze,Dupe2,Hagemeister16,Hagemeister}. The real spin structures of these skyrmions naturally differ from those derived from  the continuous  model based on classical physics.

For the above reasons, we carry out simulations here for a Fe$_{0.5}$Co$_{0.5}$Si-like thin film  by means of  a quantum
computational method which  we develop in recent years
\cite{liujpcm,liupssb,LiuIan16,LiuIan17,LiuIan18,LiuIan19,LiuIan19SM}.
 Following the similar methodology \cite{Yu10}, we model the thin film as a  2D magnetic   monolayer, and impose the  periodic boundary conditions to mimic the quasi-infinite size   of the system.
  In brief, the phenomena observed in our simulations can be summarized as below: In the absence of external magnetic field,  the helical texture is the   ground   state. \emph{When}  a weak magnetic field   is applied  normal to the   monolayer, SkX textures of   hexagonal-closely-packed (HCP)  pattern are   induced \cite{Yu10}, and  the  periodic distances  for the helical and SkX states  agree roughly with   the theoretical values \cite{Yu10,Leonov}. As the field strength is enhanced  and temperature  changes, the SkX textures can be  rotated or/and  deformed, and their periodicitic distances   increases. When  the external field is sufficiently strong, a skyrmion plus  bimeron  phase appears, which \emph{still looks} symmetric  geometrically.  Within   further increased external magnetic field, the magnetic system finally becomes ferromagnetic as expected.   Very interestingly,  the  calculated topological charge density for every SkX  also forms periodical and symmetric lattice which is almost identical to the corresponding magnetic SkX,  demonstrating the correctness of our simulations.  This quantum model, when scaled, is able to qualitatively reproduce the experimental results for SkX of large spacing distance. On the other hand, when the skyrmion size is around a few nano-meters in diameter, or more general,\emph{ as the spin-spin  correlation length of the material is in the scale of lattice constan}t, the quantized discrete model is expected to be more accurate than the classical ones.

\section{The Quantum Computational Method and Related Theory}

 To  simplify the model,  a quasi-2D system like Fe$_{0.5}$Co$_{0.5}$Si thin  film can be modelled  as a monolayer  ferromagnet with the square crystal structure \cite{Yu10}, and its Hamiltonian be expressed as
\begin{eqnarray}
{\cal H} = & -\frac{1}{2}\sum_{<i,j>}\left[{\cal J}_{ij}{
\vec{S}_i \cdot }\vec{S}_j
- D_{ij}{\vec{r}_{ij}\cdot(\vec{S}_i\times}\vec{S}_j)\right]\nonumber\\
\space & -K_A\sum_i\left(\vec{S}_i\cdot \hat{n}  \right)^2
-\mu_Bg_S{\vec B }\cdot\sum_i{\vec S}_i \;.
 \label{hamil}
\end{eqnarray}
Here, the first two terms represent the Heisenberg exchange and DM
interactions with the strengths of ${\cal J}_{ij}$ and $D_{ij}$
between a pair of spins at the $i$- and $j$-th lattice sites,
respectively, and $ <ij>$  means that these interactions are
limited between the nearest neighboring spins; the third term
stands for the uniaxial anisotropy assumed
 to be perpendicular to the monolayer plane,  and the last one denotes the Zeeman
 energy of the 2D  system within an applied magnetic field.

The  quantum computational  method employed in the present work
has been described in details in our published papers
\cite{liujpcm,liupssb,LiuIan16,LiuIan17,LiuIan18,LiuIan19,LiuIan19SM}.
So the spins appearing in the above Hamiltonian  are  quantum operators instead of classical vectors, and all physical quantities are calculated with quantum formulas. In the Heisenberg representation, when $S$ = 1  for example  as it is assumed in our simulations, the matrices  of the three spin components are:
{\footnotesize
\begin{eqnarray}
S_x = &\frac{1}{2}\left(
\begin{array}{ccc}
0 &  \sqrt{2}  & 0\\
\sqrt{2} & 0 & \sqrt{2}  \\
0 & \sqrt{2}  & 0\\
\end{array}
\right)\;, & S_y =\frac{1}{2i}\left(
\begin{array}{ccc}
0 &  \sqrt{2}  & 0\\
-\sqrt{2} & 0 & -\sqrt{2}  \\
0 & \sqrt{2}  & 0\\
\end{array}
\right)\;,\\ \nonumber
 S_z = & \left(
\begin{array}{ccc}
1 &  0  & 0\\
0 & 0 & 0 \\
0 & 0  & -1\\
\end{array}
\right)\;,  & \space\; \\ \nonumber
\end{eqnarray}\nonumber
}
respectively.

In light of molecular field theory, the $i$-th spin is considered to be under the interaction of an effective magnetic  field ${\bf B}_i^M$ generated by the  neighboring  magnetic moments. As a result, if the lattice chosen for simulations consist of $N$ spins,  the Hamiltonian of the whole magnetic system shown  above can be decomposed into $N$ coupled Hamiltonians.  Each of them is for a spin, and that for the $i$-th spin   is given by
\begin{eqnarray}
{\cal H}_i = & -\sum_{<i,j>}\left[{\cal J}_{ij}{
\vec{S}_i \cdot }\langle\vec{S}_j\rangle
- D_{ij}{\vec{r}_{ij}\cdot(\vec{S}_i\times}\langle\vec{S}_j\rangle)
\right]\nonumber\\
\space & -K_A\sum_i\left(\vec{S}_i\cdot \hat{n}  \right)^2
-\mu_Bg_S{\vec B }\cdot\sum_i{\vec S}_i \;.
 \label{shamil}
\end{eqnarray}
 This   Hamiltonian of a single-spin  can be easily diagonalized,   and  the thermal average of any physical variable $A$ at temperature $T$ can be   calculated with
\begin{equation}
\langle A\rangle = \frac{{\rm Tr}\left[\hat{A}\exp(\beta{\cal H}_i)\right]}{{\rm Tr}\left[\exp(\beta{\cal H}_i)\right]}\;,
\label{avq}
\end{equation}
where    $\beta = -1/k_BT$.

 All of our recent simulations  are started from a random magnetic configuration  above the  transition temperature, then carried out stepwise down to very low temperatures with a  reducing temperature step $\Delta T < 0$.
This trick  is very important, since at high temperatures,  the effective magnetic field is relatively weak,   the thermal interaction is  strong enough to help the spins overcome the energy barriers,  so that the code can avoid being trapped in local energy minima, and finally converge down to correct equilibrium states spontaneously.  Obviously, for this purpose, $|\Delta T|$ cannot be too large.

  At a given  temperature, the spins are  selected  one by one successively to evaluate their thermally averaged  values.  After all spins in the sample have been visited once, an iteration is completed. The iterations are repeated in a self-consistent manner. When the ratio $(|\langle \vec{S}'_i\rangle - \langle \vec{S}_i\rangle|)/|\langle \vec{S}_i\rangle|$ between two successive iterations for every spin in the lattice is less than a very small given value $\tau_0$, convergency is considered to be reached.

The   periodical length $\lambda_0$ of the SkX observed in the
Fe$_{0.5}$Co$_{0.5}$Si film plane is around 90 nm \cite{Yu10}, and
the lattice constant $a$ is about 0.45 nm \cite{Kwon13}. Thus,
the nearest pair of skyrmions are about 200 $a$ apart. In order to
display the calculated SkX texture for the Fe$_{0.5}$Co$_{0.5}$Si
film, the spin wavelength $\lambda$ was assigned to 10 by Yu et
al.~in their Monte Carlo simulations \cite{Yu10}. In this scaled
model, $\lambda$  is measured in the unit of the side-length of a
grid,   and each grid contains $n\times n$ spins where $n =
\lambda_0/(\lambda a) $ if both $\lambda_0/a$ and $\lambda$ are
much larger than 1 \cite{Kwon12}. Moreover, the periodic distance
of the chiral texture can be determined by the relative  strengths
of Heisenberg exchange and DM interactions \cite{Yu10},
\begin{equation}
\tan\left(\frac{2\pi}{\lambda}\right) = \frac{D}{\sqrt{2}{\cal
J}}\;. \label{avelength}
\end{equation}
So, once  $\lambda$ is given, the ratio $D/{\cal J}$ can be estimated with this formula.

For a discrete spin model, the winding number density of a skyrmion at the $i$-th site  can be expressed as
\begin{equation}
\rho_i =
\frac{1}{16\pi}\vec{S}_i\cdot\left[(\vec{S}_{i+\hat{x}}-\vec{S}_{i-\hat{x}})\times
(\vec{S}_{i+\hat{y}}-\vec{S}_{i-\hat{y}}) \right]\;. \label{rhoQ}
\end{equation}
Here an additional  prefactor $1/4$ has been \emph{introduced},  since
the spacial distances between  the spin pair $
(\vec{S}_{i+\hat{x}}$, $\vec{S}_{i-\hat{x}}) $ in the $x$-direction,  and another pair $(\vec{S}_{i+\hat{y}}$, $\vec{S}_{i-\hat{y}}) $  in the $y$-direction, are both equal to 2$a$. Moreover, all spins must be normalized before being inserted to this formula. Afterwards, the averaged winding number per skyrmion can be estimated by dividing the sum over the lattice, $\sum_i\rho_i$, with the total number of skyrmions observed in the lattice.  To
check the  correctness of above formula,  the winding numbers have also been   calculated with  a formula employed by previous authors \cite{JMLiu}, which gives the same results.

On the other hand, the helicity of a chiral spin texture is defined as \cite{JMLiu,Shibata}
\begin{equation}
\gamma
=\frac{1}{N}\sum_i\left(\vec{S}_i\times\vec{S}_{i+\hat{x}}\cdot\hat{x}
+\vec{S}_i\times\vec{S}_{i+\hat{y}}\cdot\hat{y} \right)\;.
\label{helicity}
\end{equation}
This quantity is attributed to the relativistic spin-orbital coupling, its sign and the magnitude  reflect the swirling direction (right-handed or left-handed), and  degree of the  chirality.

\section{Computational Results}

When the spin wavelength  $\lambda$ is assigned to 10, the ratio $D/{\cal J}$ is  determined to be 1.027 from Eq.(\ref{avelength}). This value is used in   all our simulations that are performed on  a 30$\times$30 square lattice, and  the periodical boundary conditions are imposed to mimic the quasi-infinity of  the 2D magnetic system. For simplicity, we further assign ${\cal J}/k_B$ = 1 K. That is, all physical quantities are scaled with the Heisenberg exchange strength ${\cal J}$ and $k_B$.

\subsection{The Ground Helical State in the Absence
 of External Magnetic Field}

 We notice from our calculated results that,  in the absence of external magnetic
field, the helical texture  is spontaneously formed below the transition  temperature, $T_M$ = 3.3 ${\cal J}$.   In Figure 1,   the $xy$ component  and $z$-contour of the calculated spin structure are both projected onto the film plane.
\begin{figure}[htb]
\centerline{
\epsfig{file=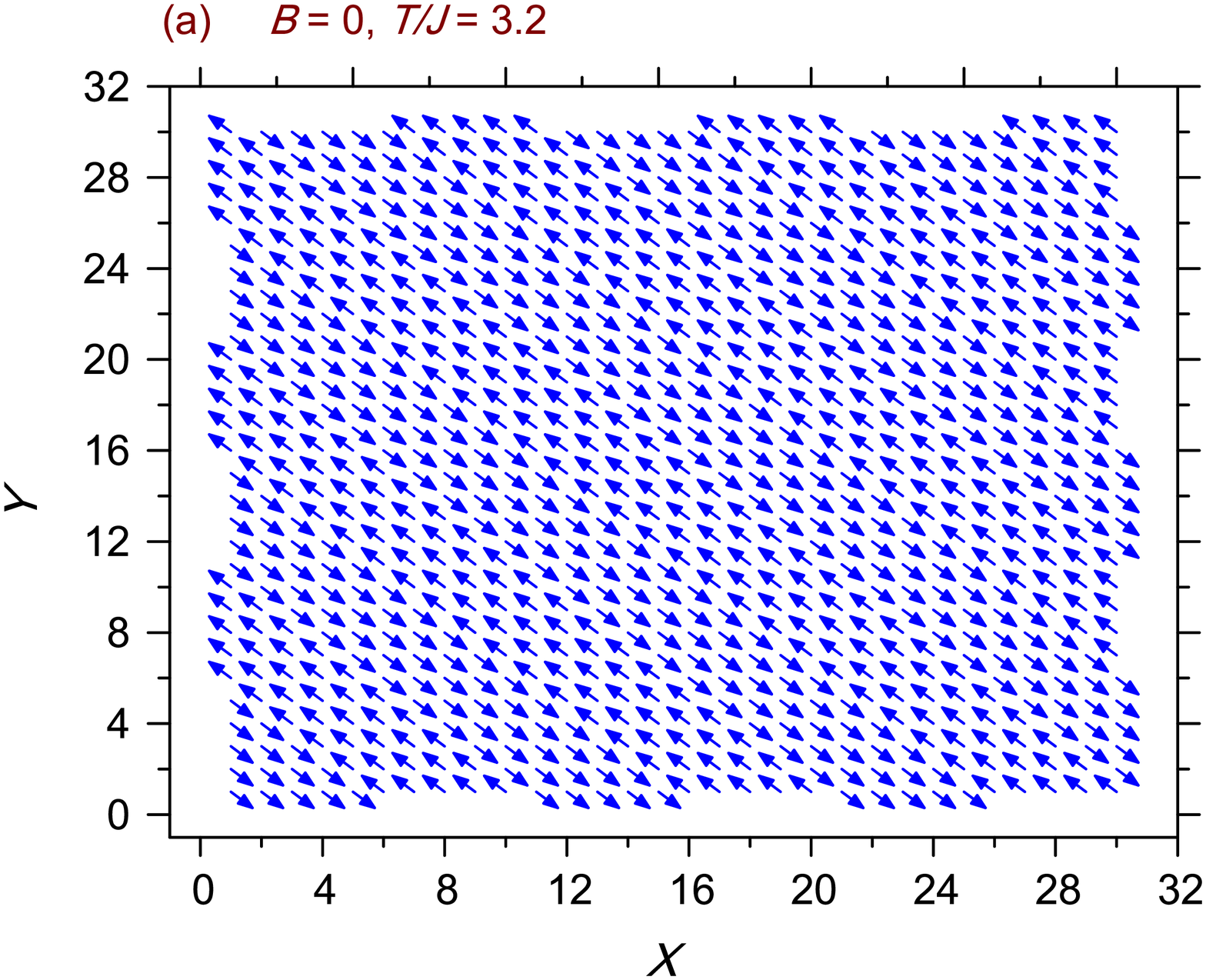,width=4.3cm,height=3.8cm,clip=}
\epsfig{file=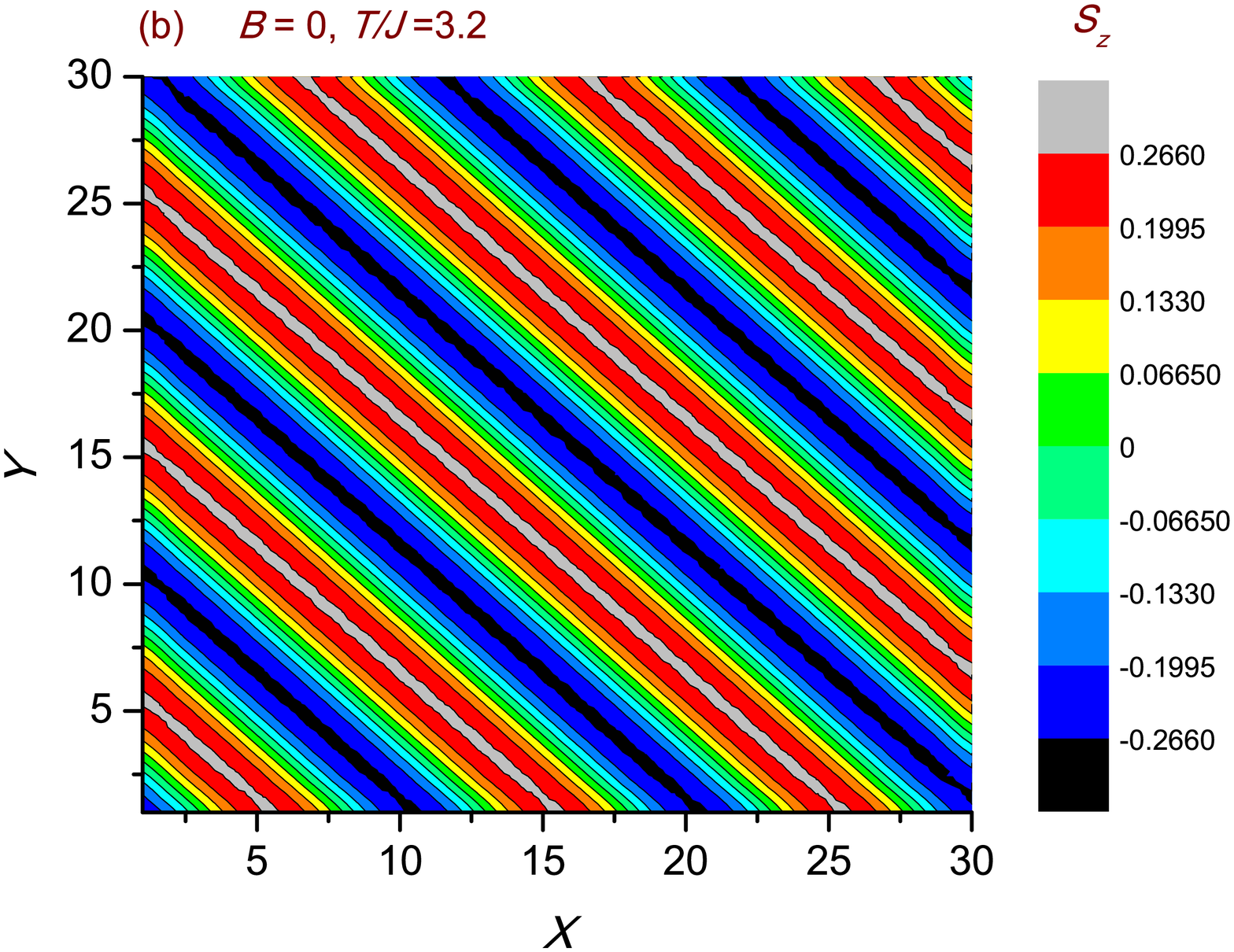,width=4.3cm,height=3.8cm,clip=}
 }
\begin{center}
\parbox{13.5cm}{\small{{\bf Figure 1.} The $xy$   texture (a) and $z$-contour (b) of the helical spin structure calculated at $T/{\cal J}$ = 3.2 in the absence of external magnetic field.
 }}
\end{center}
\end{figure}
 It is interesting to find there that the periodical lengths  in   the $x$ and $y$ directions of the helix state are all equal to 10 in the unit of grid side length as theoretically predicted. In every spatial period, there exist two spin strips which are anti-parallel in the [-110] direction on the $xy$-plane, and across each period, the $z$-components of the spins change gradually from the minimum with the most negative value to the maximum,  then  falls  down  gradually. This helical
texture remains unchanged down to almost zero temperature, but the magnitudes of all three components   increase   with decreasing temperature   as expected.

 \subsection{Initial SkX Textures Induced by a Weak External Magnetic Field}

When an external magnetic field is applied  perpendicular  to the monolayer plane,  the helical structure persists until the field is increased  to 0.1 Tesla.
\begin{figure}[htb]
\centerline{
 \epsfig{file=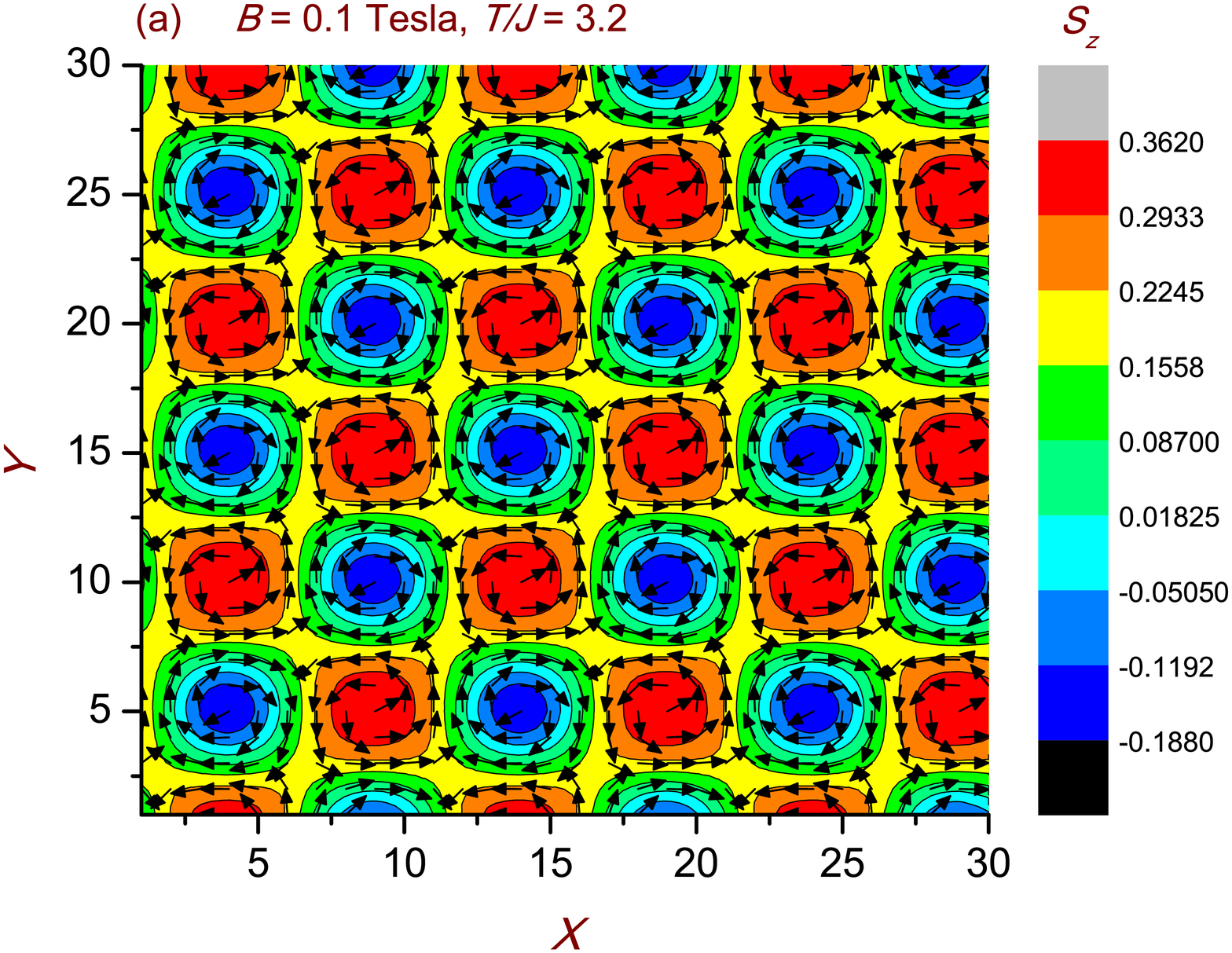,width=4.3cm,height=3.8cm,clip=}
 \epsfig{file=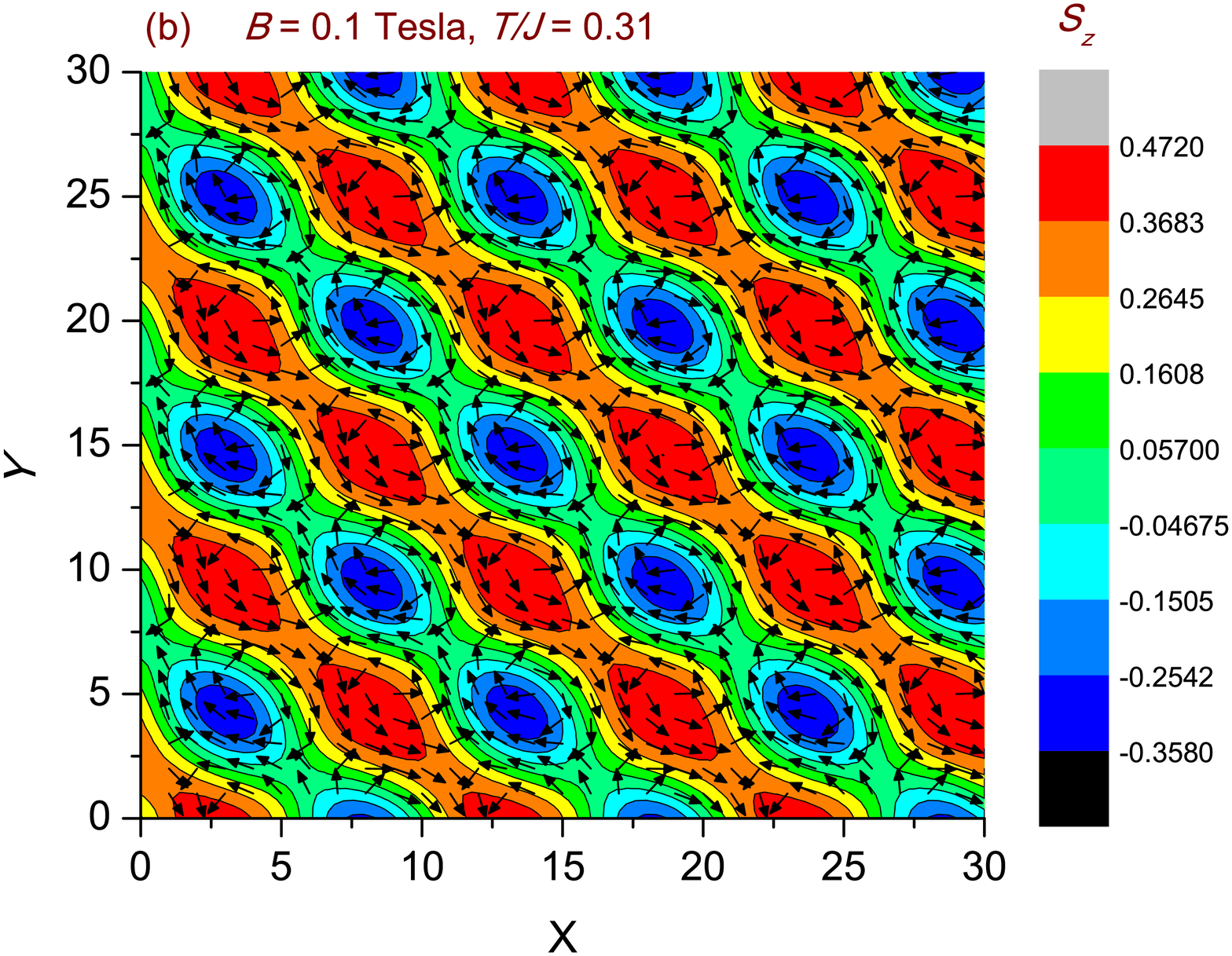,width=4.3cm,height=3.8cm,clip=}
 }
\begin{center}
\parbox{13.5cm}{\small{{\bf Figure 2.} The
SkX textures  calculated at (a) $T/{\cal J}$ = 3.2, and (b)
$T/{\cal J}$ = 3.1  in an external magnetic field of 0.1 T applied
normal to the monolayer plane. }}
\end{center}
\end{figure}
However, within the magnetic field of this strength, magnetic SkX can be induced  at $T/{\cal J}$ = 3.2, which is immediately  below the ferromagnetic phase, as displayed in Figure 2(a). There, 18 skyrmions (printed in blue) are observed on the
monolayer plane in a pattern  of hexagonal-close-packed (HCP) crystal structure, and they are all separated by  shallow vortices (displayed in red). The two sides of such a hexagon are parallel with the $x$-axis, so it is  referred  as 'regular'  to
distinguish it from those rotated and deformed  hexagons. However, the external magnetic field is now too weak to stabilize the SkX structure. When temperature falls down to $T/{\cal J}$ = 0.31, the skyrmions and vortices are all deformed, they are elongated in the [-110] direction, but the chiral spin texture still keeps the
regular HCP pattern as shown in Figure 2(b). As the temperature drops  further, the SkX texture disappears, being replaced by the helical structure  afterwards.   \emph{Hou et al.  have reported the similar phenomenon before \cite{Hou}. In their  Monte Carlo simulations done for a 2D  chiral magnet,  a  surprising upturn of the topological charge was identified at  high temperatures. They attributed this upturn to spin fluctuations, that is,  the topology was  believed to be thermally induced.}

\subsection{Typical SkX State Induced by Moderate External Magnetic Field}
As  the external magnetic field is further  increased, a stable SkX phase appears over a broad temperature range. For instance, when $B_z$ = 0.11 T, SkX is observed in the low temperature region $T/{\cal J} \leq$ 3.2. Figure 3(a,b)  show the SkX textures calculated at $T/{\cal J}$ = 3.2 and 0.1  respectively   within
this field.
\begin{figure}[ht]
\centerline{
 \epsfig{file=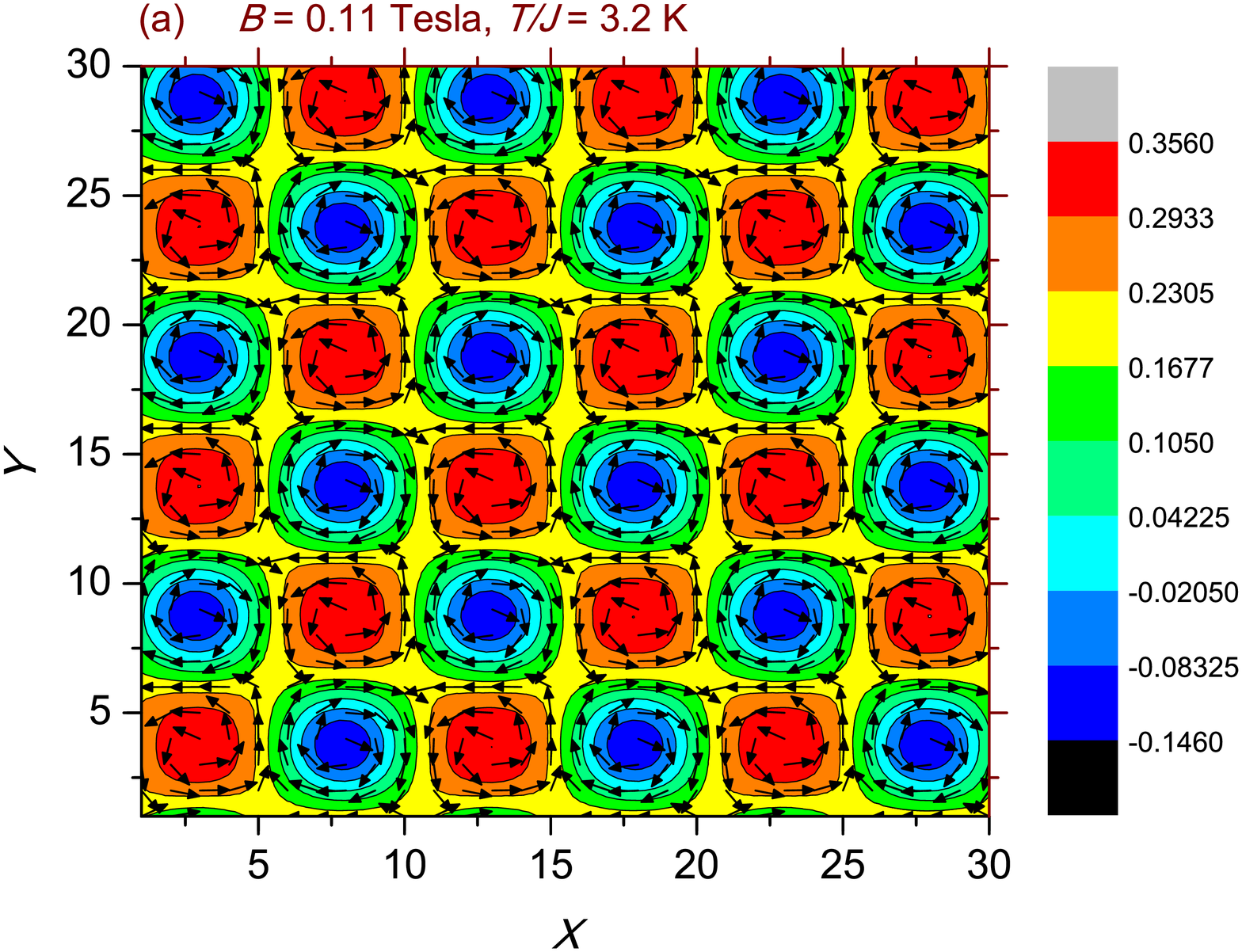,width=4.3cm,height=3.8cm,clip=}
 \epsfig{file=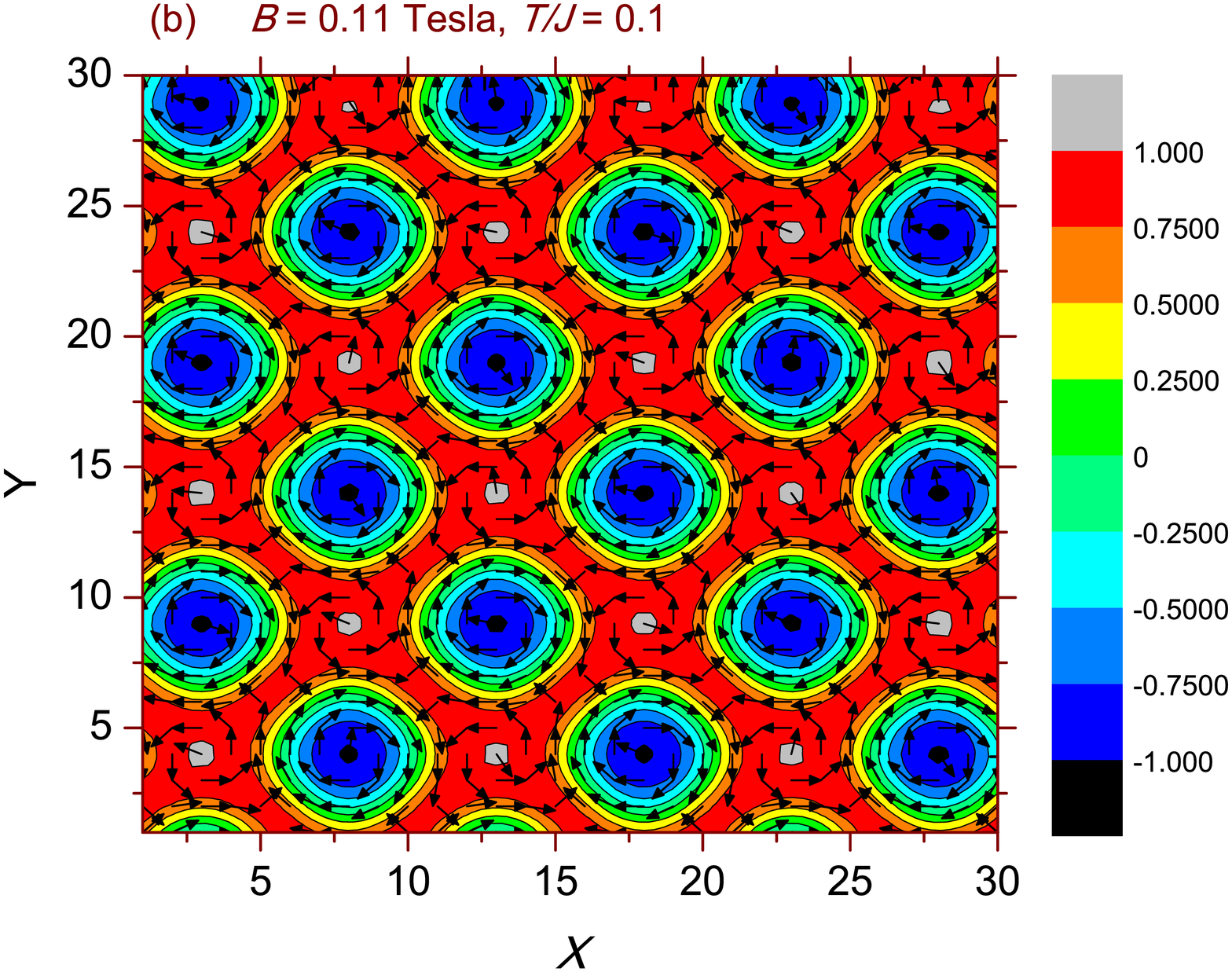,width=4.3cm,height=3.8cm,clip=}
 }
\begin{center}
 \parbox{13.5cm}
 {\small{{\bf Figure 3.}
The SkX textures  calculated at (a) $T/{\cal J}$ = 3.2, and (b)
$T/{\cal J}$ = 0.1,  within an external magnetic field of 0.11 T
applied perpendicular to the monolayer plane.
  }}
 \end{center}
\end{figure}
There, every skyrmion is surrounded by four non-fully developed vortices.   Since $D$ is positive, these   two sorts of spin textures are both right-handed. They appear alternatively in the lattice plane. All skyrmions curl clockwise, while the
vortices swirl anti-clockwise, so as to minimize the total (free) energy of the whole system.

\emph{However, in Ref.\cite{Yu10}, no  surrounding vortices  were observed in experimental results and classical Monte Carlo (CMC) simulations. The reasons are given below. To simulate the Fe$_{0.5}$Co$_{0.5}$Si-Like 2D system, a scaled model is  used here. In this model,  one spin represent several hundreds of lattice sites. Thus, vortices curling in the opposite direction must appear  to reduce the sudden energy increase caused by the closely packed skyrmions. However, in real Fe$_{0.5}$Co$_{0.5}$Si film, the skyrmions  are  actually quite large, so the spins can change their directions continuously, thus much smaller or no  surrounding vortices  are required for the spin texture to relax. On the other hand, in the interstitial  areas between the skyrmions, the spins have been considerably rotated by the  external magnetic field to the out-of- plane direction, whereas the on plane components are greatly reduced. Consequently, even the vortices appear in the interstitial areas, they are hardly to be observed in experiments. In contrast, before our simulated spin textures are plotted, the on-plane components of all spins have been normalized, so that the vortices can be easily seen. In Figure 1(b) of Ref.\cite{Yu10}, we can see that interstitial areas between the skyrmions are quite large.   However, the vortices are missing in these  areas. The SkX shown there was obtained in their CMC simulations. We guess that the authors did not do  the normalization as we do here.  }

As expected, though the external magnetic field is applied along the $z$-axis, the spins inside the skyrmion cores align in the opposite direction. \emph{The increase in total energy incurred by these spins  are offset by the surrounding vortices, where the spins have been greatly rotated by the external field toward its orientation.}

Figure 2(a) and Figure 3(a) are obtained at the same temperature, they look very similar at the first glance.  However, the vortices and skyrmions have
actually exchanged their positions within  external magnetic field of different strengths. Moreover, by comparing Figure 3(a) and (b),  it is observed  that as temperature drops down, the cores of skyrmions expand,  but those of the vortices shrink spatially. This rule in general holds   in all later cases.

\subsection{Vortex Lattice and SkX Formed  at $B_z$ = 0.13 T}
When   $B_z$ =  0.13 T, an extremely  shallow vortex lattice is
observed at $T/{\cal J}$ = 3.2 immediately below the transition
temperature as displayed in Figure 4(a).
\begin{figure}[htb]
 \centerline{
\epsfig{file=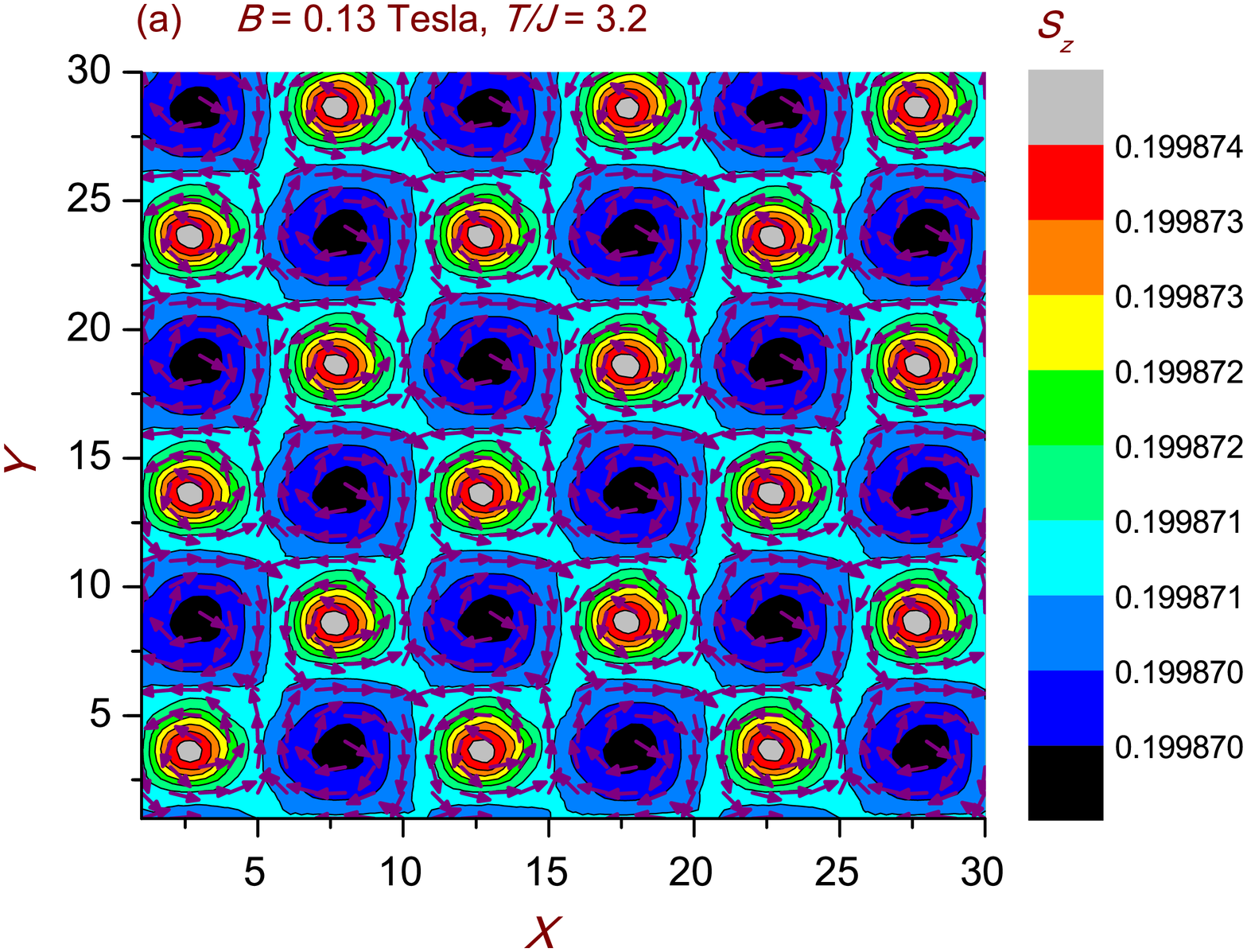,width=4.3cm,height=3.8cm,clip=}
\epsfig{file=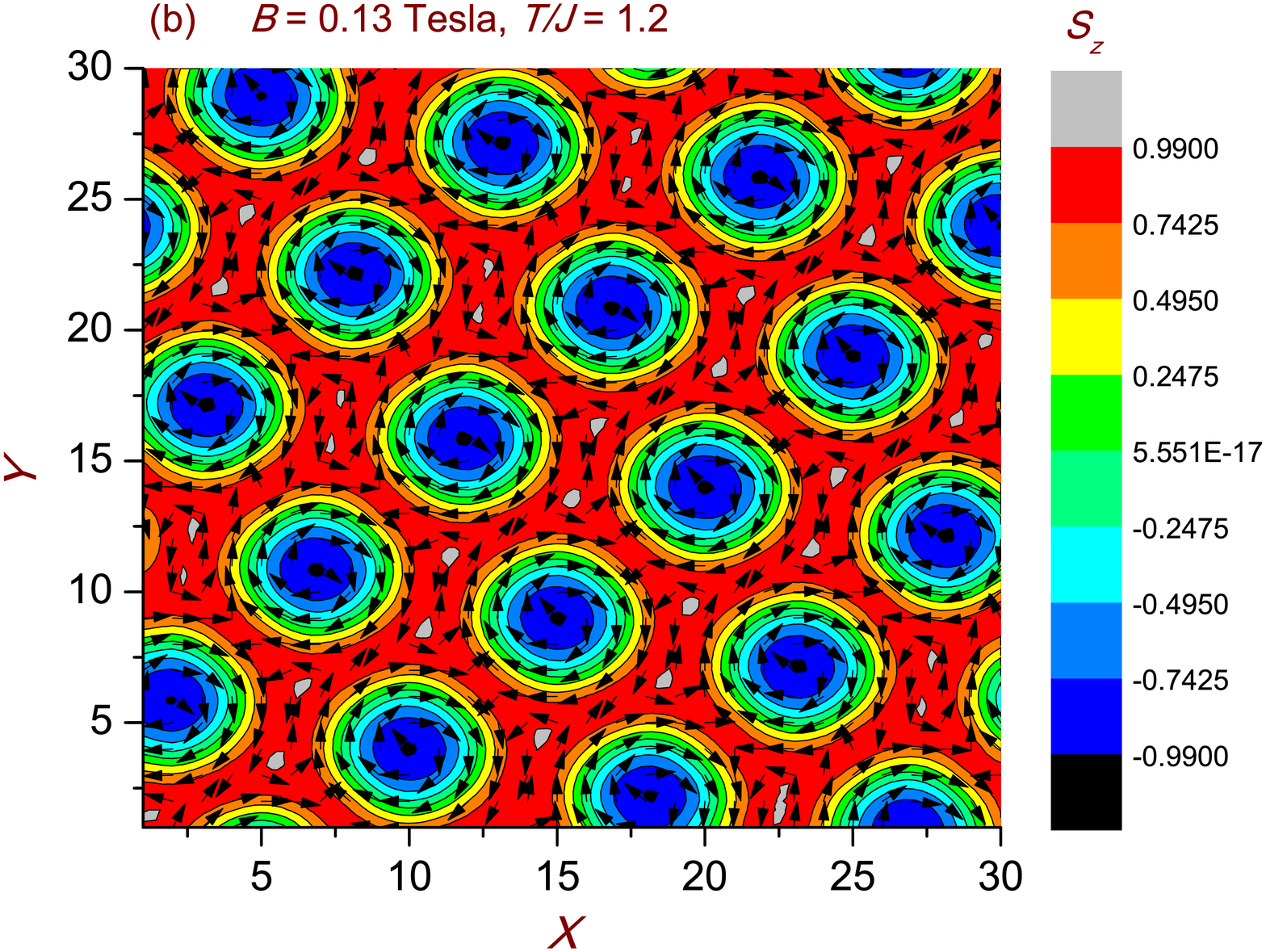,width=4.3cm,height=3.8cm,clip=}
 }
\begin{center}
 \parbox{13.5cm}
 {\small{{\bf Figure 4.}
The (a) vortex lattice at $T/{\cal J}$ = 3.2, and (b) SkX texture at $T/{\cal J}$ = 1.2,  induced by external magnetic field of 0.13 T  perpendicular to the monolayer plane. }}
 \end{center}
\end{figure}
 The $xy$ projection of this  chiral lattice pattern  looks  similar to  that depicted in Figure 2(a): 18 vortices curl clockwise,  other 18 vortices  swirl anti-clockwise, and the two sets of vortices appear alternatively on the $xy$-plane. It can   be inferred   that such  shallow vortical lattices may also be induced immediately below FM phase by external magnetic field of other  strengths, but they are  easily destroyed by thermal disturbance, thus can hardly be observed in experiments. When $T/{\cal J}$ drops to 3.1, the regular HCP SkX appears, and maintains until $T/{\cal J} $ =  1.3  over a wide temperature range. While $T/{\cal J} \leq$   1.2, the SkX is found to be rotated clockwise for 6$^{\rm o}$ by the effective magnetic field. However, the whole spin structure still keeps  HCP pattern as depicted  in Figure 4(b).

 \emph{The spin texture rotation just described above  seems very strange, readers may wonder what are the reasons behind. Firstly, the self-consistent algorithm has been implemented into our computing  program, so the  code can converge spontaneously to the equilibrium states. That is,  all spin textures presented in the paper are self-organized, the rotations can not be phantom  simply caused by  artificial intervention.}

 \emph{Such phenomena will be observed in following figures as well. Theoretically speaking, as temperature or/and the strength of magnetic field vary, not only  the magnitudes, but also the orientations of all spins change correspondingly. Since the spins in the system are correlated directly or indirectly, the process may inevitably give rise to spin texture rotation to  fit new magnetic configuration. On the other hand, we will see latterly  that enhanced magnetic field usually leads to enlarged periodicity. This  may also result in SkX rotation so that the spins can adjust themselves to meet the requirements of the new symmetry and fit the periodical boundary conditions.}

 \subsection{Evolution of   SkX Texture Driven by Enhanced External Magnetic Field}

At $B_z$ = 0.15 T,  the same   phenomena just described  are observed once again. In the temperature range 3.2  $ > T/{\cal J} >$ 1.9, the spin texture is  a SkX  of regular HCP pattern. However, as temperature falls  below $T/{\cal J}=$ 1.9, the SkX texture is  not only rotated, but also  deformed as shown in Figure 5(a).

\begin{figure}[htb]
\centerline{
 \epsfig{file=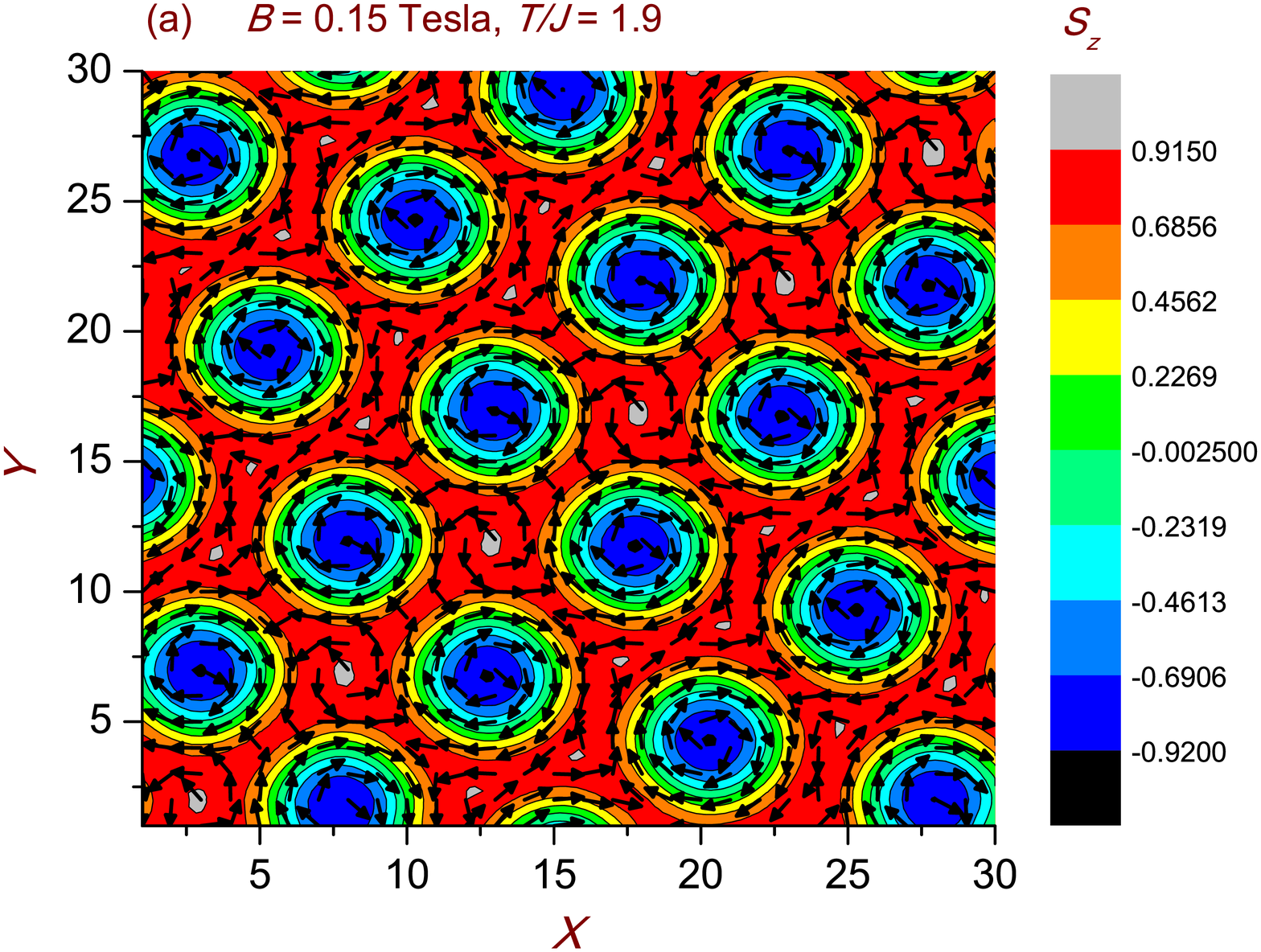,width=4.3cm,height=3.8cm,clip=}
 \epsfig{file=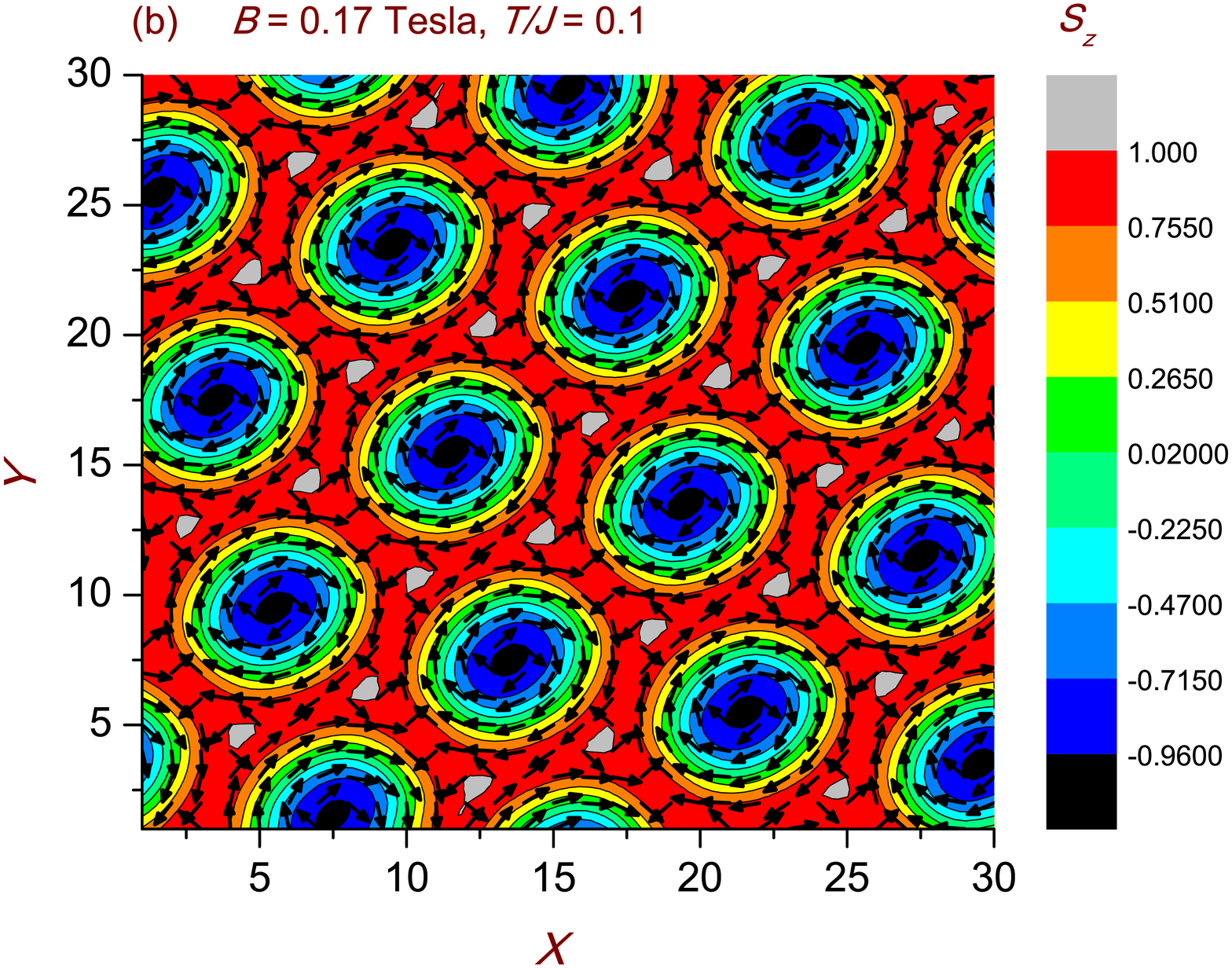,width=4.3cm,height=3.8cm,clip=}}
 \centerline{
 \epsfig{file=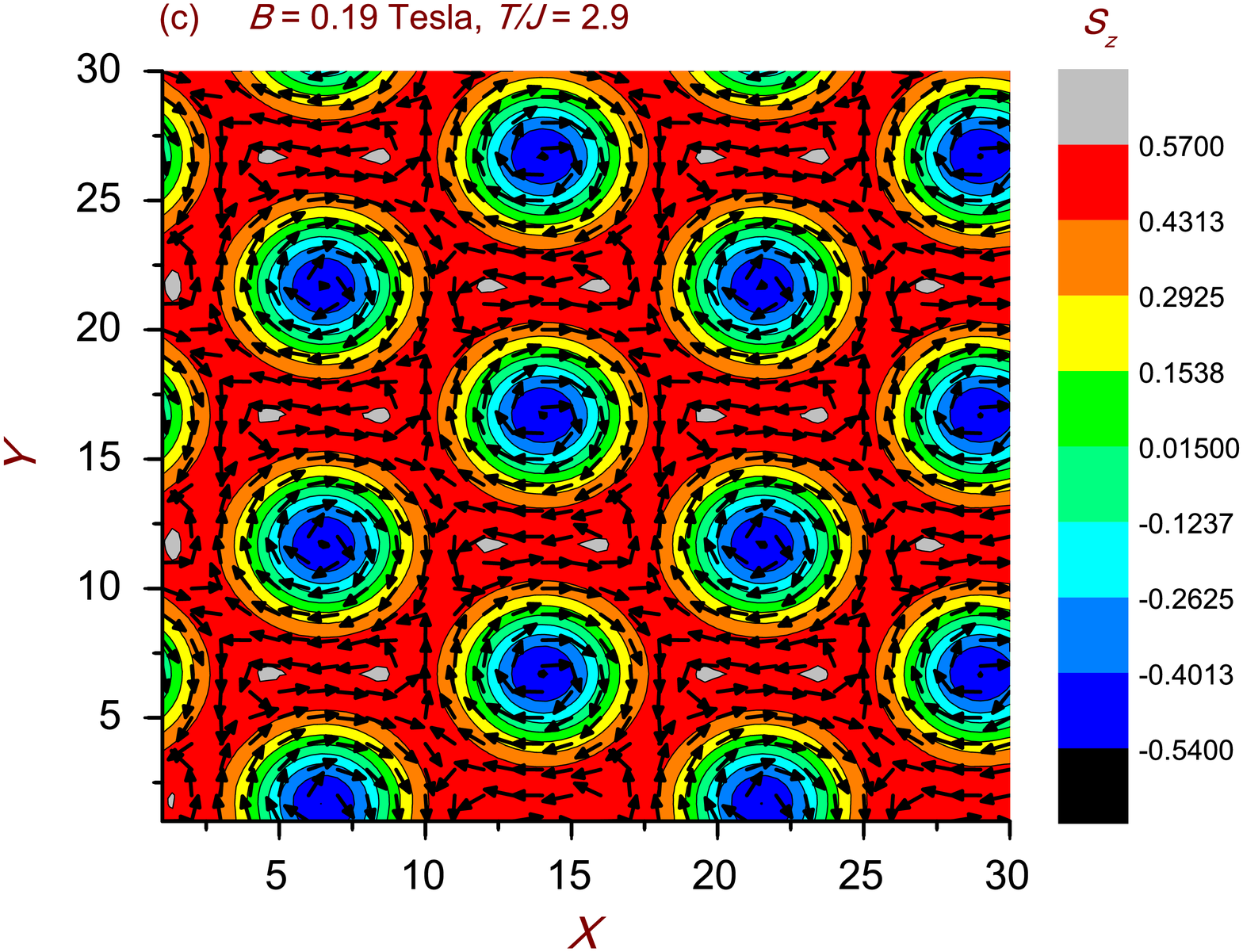,width=4.3cm,height=3.8cm,clip=}
\epsfig{file=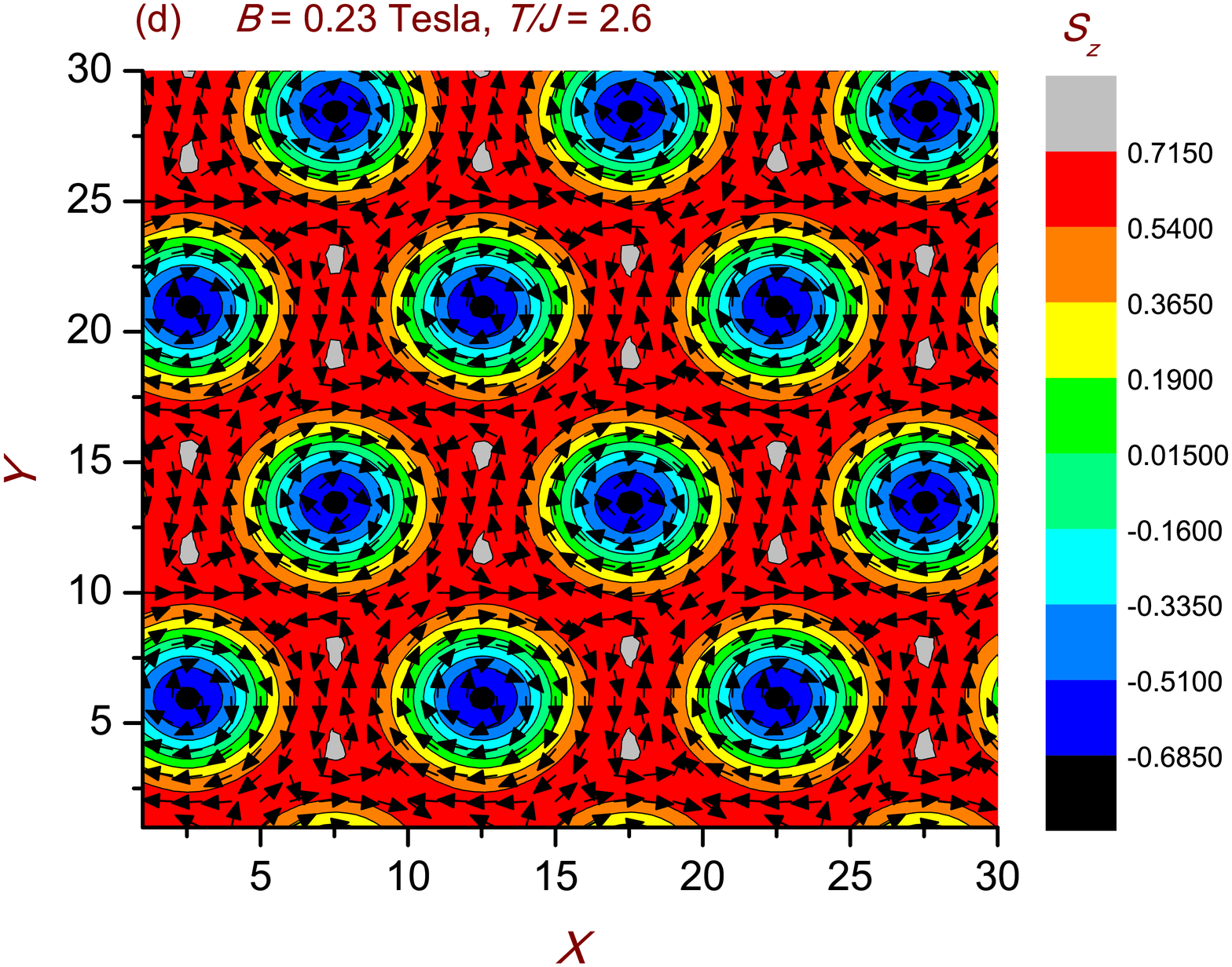,width=4.3cm,height=3.8cm,clip=}
 }
\begin{center}
 \parbox{13.5cm }
 {\small{{\bf Figure 5.}
The  skyrmion lattices simulated when (a) $B$ = 0.15 T, $T/{\cal
J}$ = 1.9, (b) $B$ = 0.17 T,  $T/{\cal J}$ = 0.1, (c) $B$ = 0.19
T, $T/{\cal J}$ = 2.9, and (d) $B$ = 0.23 T, $T/{\cal J}$ = 2.6,
respectively.
  }}
 \end{center}
\end{figure}

As   $B_z$ is increased to 0.17 T, the SkX is rotated by the effective magnetic field  once it is formed immediately below the polarized FM phase, and this chiral texture persists  from $T/{\cal J}$ = 3.0 down to very low temperature as shown in Figure 5(b). Now, the HCP SkX pattern is not deformed,  however,  every skyrmion has been considerably enlarged,  and the total number of skyrmions $N_s$   reduced to 15.

As depicted in Figure 5(c), when $B_z$ reaches 0.19 T, the SkX texture has been rotated clockwise further, and the total skyrmion number $N_s$   reduced to 12 in a wide temperature  below $T/{\cal J}$ =2.9.

Very interestingly, when  $B_z$ is increased to,   for examples,  0.23, 0.25  and 0.27 T,   the   spin textures are found to re-assume the regular HCP SkX pattern   in the  low temperature range as displayed in Figure 5(d), and the total skyrmion numbers are equal to 12 in all these cases.

 \subsection{Skyrmions Plus  Bimerons Texture Formed in the Transition Field}

The SkX texture can be maintained until the external magnetic field is  increased up to the critical value $B_z^c$ = 0.28 T.  At $T/{\cal J}$ = 2.4,  which is just below the FM phase,   the total skyrmion number $N_s$ is reduced to 8,  but the skyrmions are still distributed  in the regular HCP pattern  as seen in Figure
6(a).
\begin{figure}[htb]
\centerline{
 \epsfig{file=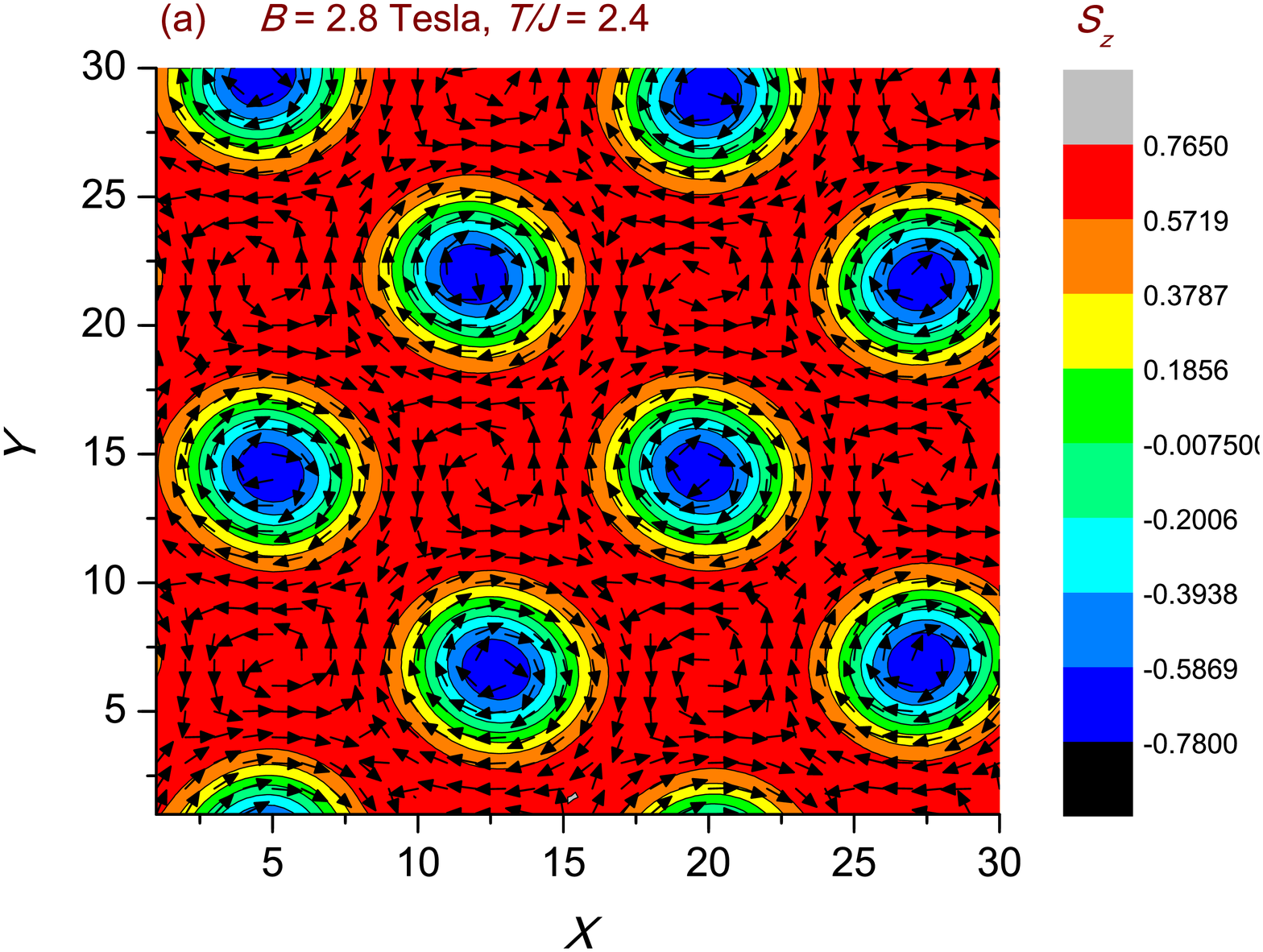,width=4.3cm,height=3.8cm,clip=}
 \epsfig{file=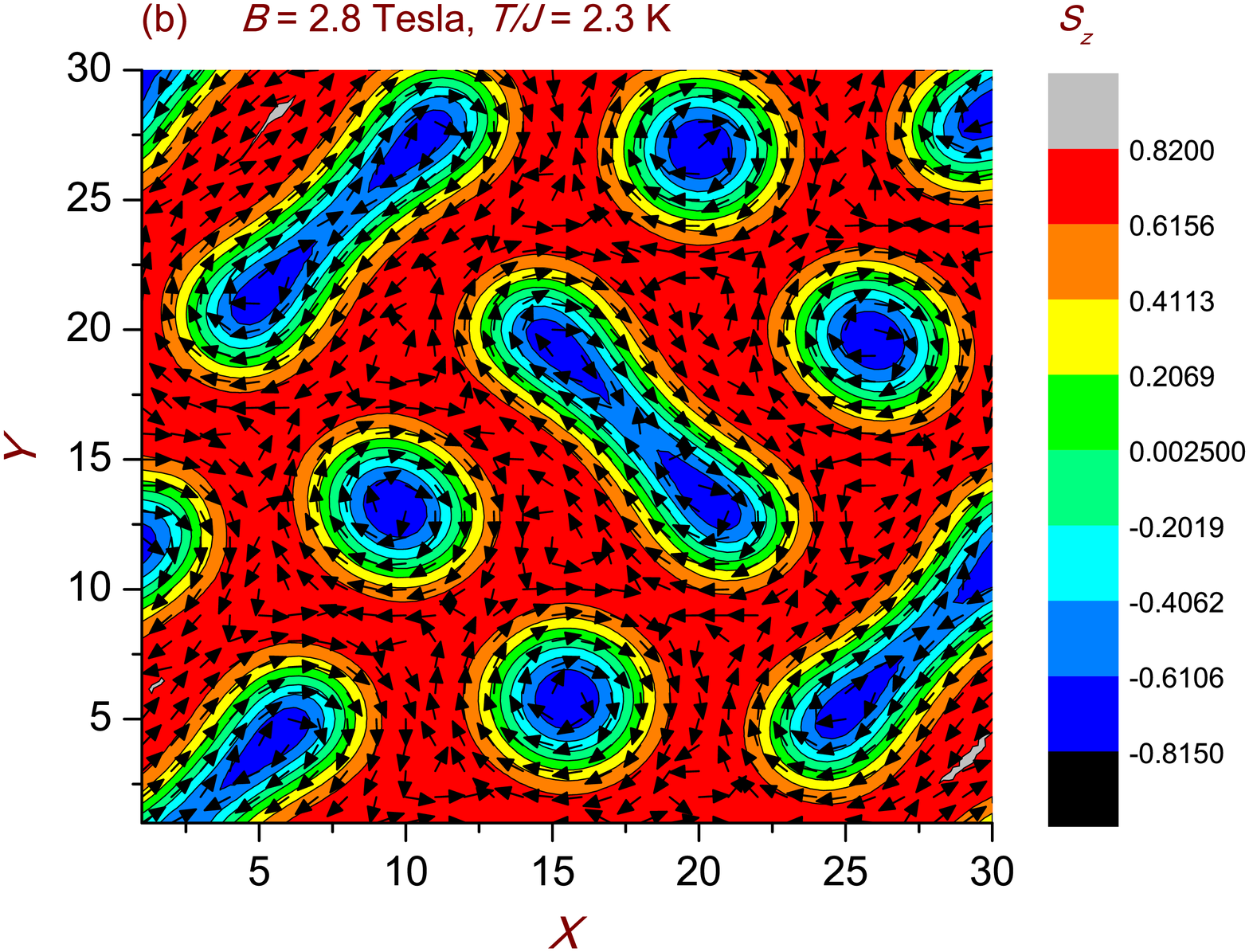,width=4.3cm,height=3.8cm,clip=}
 }
\begin{center}
 \parbox{13.5cm}
 {\small{{\bf Figure 6.}   When $B_z$ = 0.28 T,  (a) SkX,   and (b) bimeron     textures, are  formed  at $T/{\cal J}$ = 2.4  and $T/{\cal J} \le $  2.3,  respectively.
  }}
 \end{center}
\end{figure}

 Within the external magnetic field of this strength, as temperature  drops to the  range $T/{\cal J} \leq $ 2.3, a helix plus SkX texture  emerges on the monolayer as shown in Figure 6(b). These helical strips are of finite lengths. They are called as bimerons, and  expected to arise above the ground helical strip states when the external magnetic field is sufficiently strong \cite{Yu10,Ezawa2,Ezawa1}.  A bimeron has two half-skyrmions at its two ends, they are connected by a rectangular strip domain. Each of the half-skyrmions bears opposite half skyrmion in the continuous model \cite{Ezawa2,Ezawa1}. The spin
structures of the full skyrmions and bimerons shown in the figure are all right-handed. They are separated by considerably deformed and non-fully developed anti-clockwise vortices,   so as to minimize the total energy of the whole magnetic system.

 \subsection{Finite Helical Strips and Sparse Skyrmions Induced
 in Strong Magnetic Field}

When $B_z$ = 0.29 T and   $T/{\cal J} \leq$ 2.3, six helical strips are observed. They are all  parallel with [110]. Four of these strips terminate inside the $ 30\times 30$ lattice, and each of them has a half-skyrmion at its one end. So we can speculate that all the strips are  finitely long, so they are actually bimerons. The ending half skyrmions are all right-handed, so are the rectangular parts of the strips since $D >$ 0. In comparison with those shown in Figure 6(b), these strips are much longer. This effect becomes more evident when $B_z$ is increased to 0.30 T: the helical strips look much longer along the diagonal of the square lattice. Moreover, in the low temperature region $T/{\cal J} \leq$ 2, the periodical distances between the strips become 10 in the $x$-direction, but 14 in the $y$-direction.

This  finite-helical-strip  texture maintains until $B_{z}$ is increased to  0.35 T. When $B_z$ = 0.36 T, four skyrmions are observed at  $T/{\cal J}$  = 1.9, 1.8, and 1.7  inside the   square lattice. However, the skyrmions  are not evenly distributed, since now the 30$\times$30 square lattice does not fit the periodical
 textures even if they exist in reality. Nevertheless, as temperature $T/{\cal J}$ falls down below 1.6,  the magnetic system re-assumes its finite-helical-strip texture again.  These strips can be classified into two groups which are orthogonal to each other as observed in experiment \cite{Yu10}. Now $\lambda$  is   about 14 in either $x$ or $y$ direction, which mismatches with the lattice size,   so that the calculated spin configuration does not look very symmetric. Therefore, to obtain accurate calculated results, we must adjust the lattice size to find the best one which has the lowest total (free ) energy.

While $B_z$ is further increased, the spin textures consisting of sparse  deformed  skyrmions and vortices can be observed, but they are quite irregular, probably  for the sake just described. \emph{These spin textures resemble  those depicted in Figure 3(g)  of Ref.\cite{Yu10}.}

Finally, as $B_z \geq $ 0.86 T, all spins are rotated by the external magnetic field to the $z$-direction, the whole system is  polarized to be completely ferromagnetic.

\subsection{Phase Diagram, Winding Numbers and Helicities of  Spin Textures}

\emph{The main part of the phase diagram  in the low temperature range is displayed in Figure 7(a)  for the 2D magnetic system. As the magnetic field strength falls in the region  0.11  T $\leq B \leq$ 0.27  T, SkX textures are induced. Below this pahse, as  $B \leq$ 0.10 T, helical strips are formed, except  the cases when $B $ = 0.10, SkX texture appears at $T/{\cal J}$ =0.32,   then it is greatly  stretched   along the [-110] direction at $T/{\cal J}$ = 0.31. And in a  field range   0.29 T $ \leq B \leq$ 0.35 T,  the  helix plus skyrmion textures emerge above the SkX phase.  Naturally,   in a  strong magnetic field, the whole system becomes  completely  ferromagnetic. }

\emph{This phase diagram resembles qualitatively that observed in experiment \cite{Yu10}. Since our used model has been scaled and the system is simplified to be ideal, disparity with experiment is inevitable. Comparing the phase diagrams  got numerically by means of our quantum approach and  the CMC method \cite{Yu10}, we find that the thermal spin fluctuations have been considerably overestimated in CMC simulation, but underestimated in our computation.  The disparity arisen from  our method  is due to the molecular-field  theory that has been implemented in the computing program. But this difficulty can be easily overcome by introducing randomly fluctuating terms or factors in the effective magnetic field.}

With   Eq.(\ref{rhoQ},\ref{helicity}) given in Sec.II,  the sums of the winding number density, $\sum_i\rho_i$, over the whole lattice and the averaged helicities per site $\gamma$ for the   spin textures  are calculated,   the  corresponding   curves are displayed in Figure 7(b,c).
\begin{figure}[htb]
\centerline{
 \epsfig{file=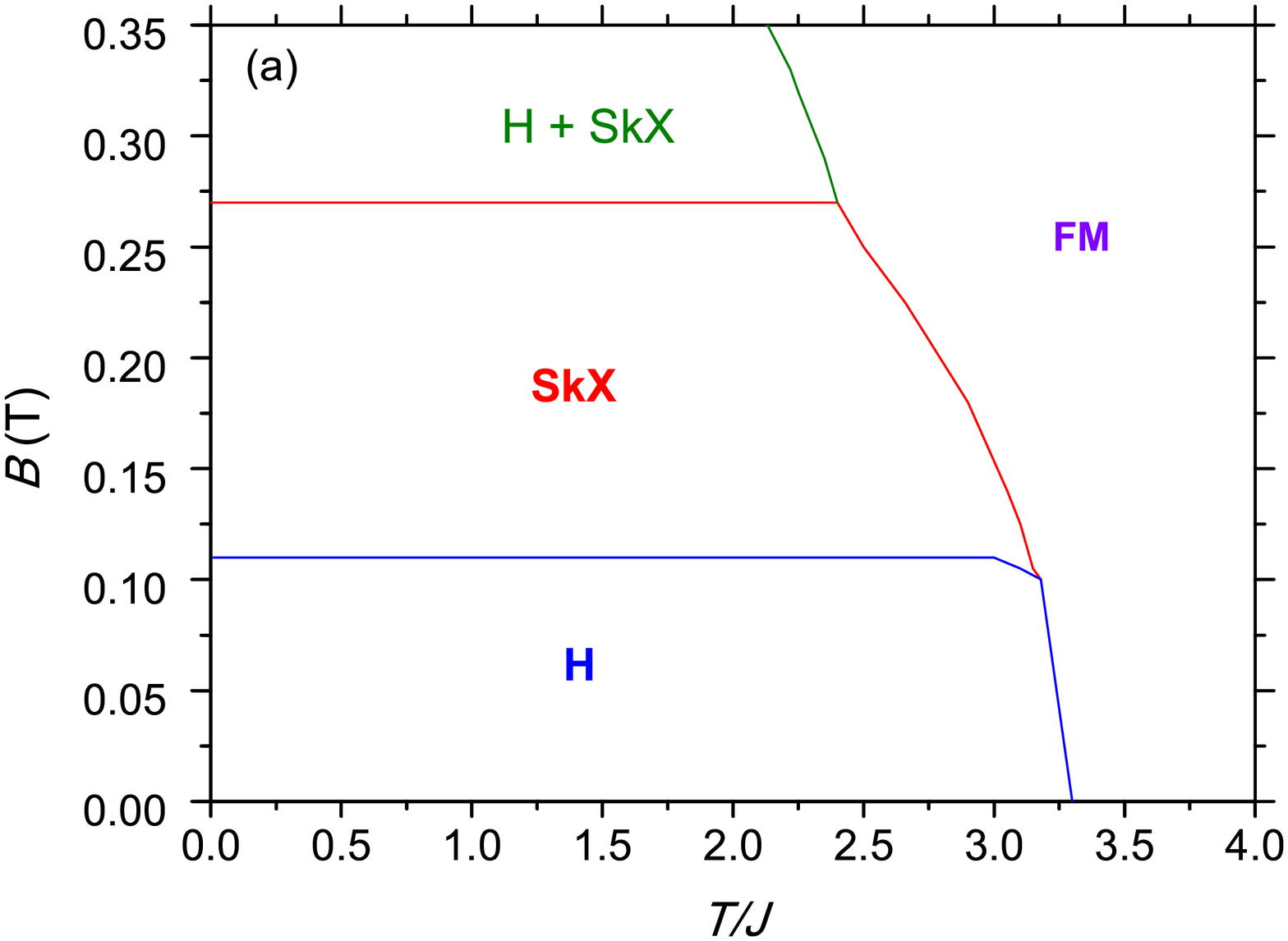,width=4.8cm,height=4.1cm,clip=}
 \epsfig{file=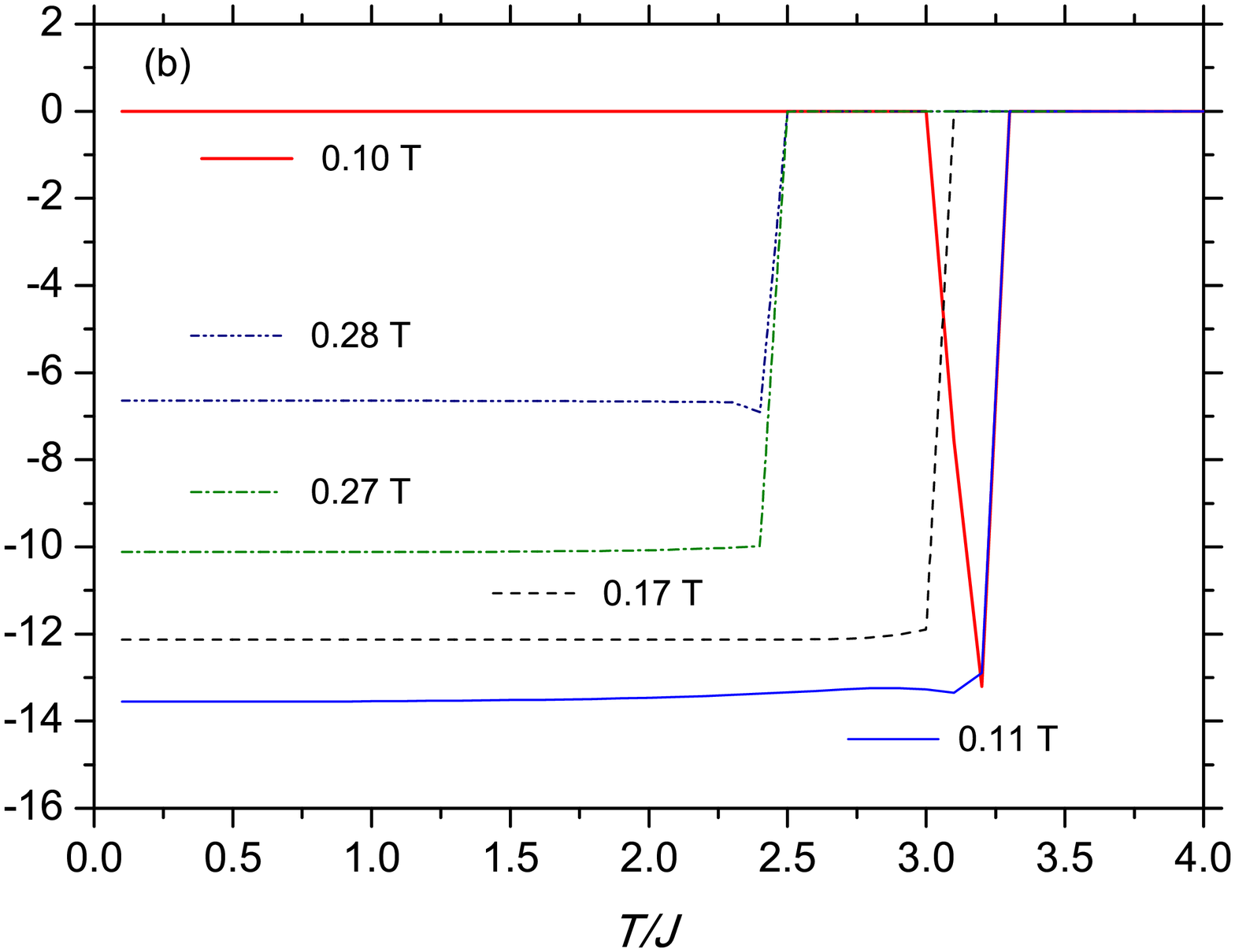,width=4.3cm,height=4.0cm,clip=}
 \epsfig{file=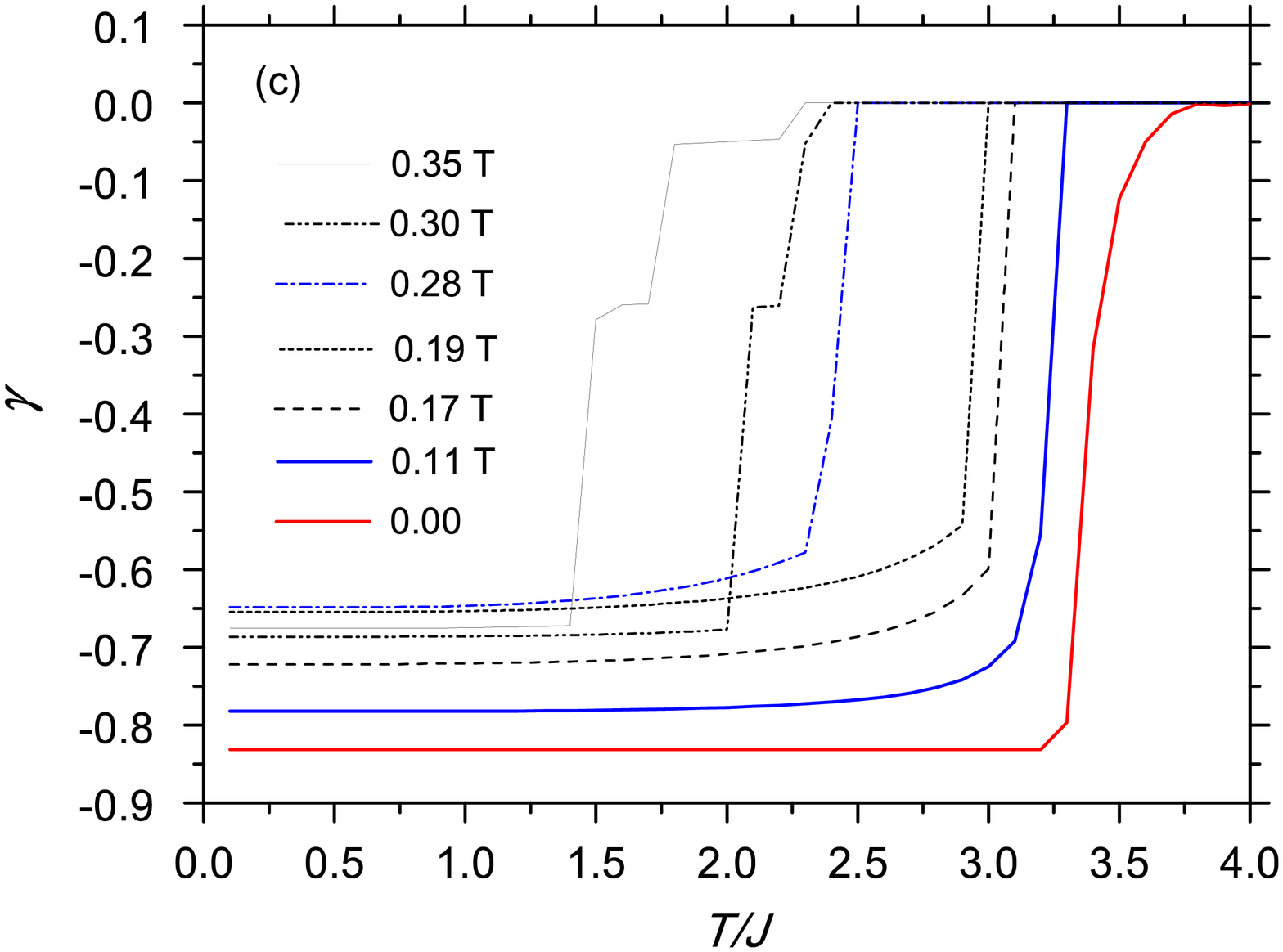,width=4.3cm,height=4.0cm,clip=}
 }
\begin{center}
 \parbox{13.5cm}
 {\small{{\bf Figure 7.}
  (a)  Phase diagram of the magnetic system, (b) total  winding number, and (c) helicity  curves of the spin textures in external magnetic fields of different strengths.  }}
 \end{center}
\end{figure}

As  shown in  above figures,  in one SkX,   a skyrmion is usually surrounded by four shallow unfully developed and frequently considerably deformed vortices of appreciable sizes. These skyrmions and their neighboring vortices are entangled together, so that it is very difficult to identify the boundaries between these two sorts of chiral spin textures. For such a SkX, the averaged winding number $Q_{av}$ per spin complex, consisting of one skyrmion and the parts of its neighboring vortices, can be estimated by dividing the sum, $\sum_i\rho_i$, with the skyrmion number $N_s$ observed inside the $30 \times 30$ square lattice. Table 1 lists these averaged numbers obtained at $T/{\cal J}$ = 0.1  in external magnetic fields of different strengths. We know already that as $B_z$ changes from 0.11 to 0.27 T, $N_s$ decreases from 18 to 12, so $Q_{av}$ is expected to change accordingly.

\begin{table}
\begin{center}
\footnotesize{
\begin{tabular}{@{}c|ccccccc}
\hline $B_z$(T) & 0.11 & 0.13& 0.15& 0.17 & 0.19 & 0.23 & 0.27\\
\hline
$N_s$ & 18 & 18 &18 & 15 & 12 & 12 & 12   \\
 $Q_{av}$ & -0.753 & -0.770 & -0.771 & -0.809  & -0.845  &
-0.845 &
-0.843 \\
\hline
\end{tabular}\\
} \caption{\small  Averaged winding numbers per
    {\it skyrmion complex} calculated at $T/{\cal J} $ = 0.1
 within external magnetic  field of different strengths.}
\end{center}
\end{table}
\normalsize

 In the absence of external magnetic field, the helical structure  is the ground state of the system, no skyrmion is formed, so   $\sum_i\rho_{i}$ = 0, and the helicity per spin $\gamma$ = -0.831 while $T/J \leq $ 3.2.

When $B_z$ = 0.10 T, the system  behaves extremely unusually. In the temperature region $T/{\cal J} \leq 3.0$, $\sum_i\rho_{i}$ is always equal to zero, that is, no skyrmion can be  formed and the helical texture dominates. However, at $T/{\cal J} = $ 3.2, $\sum_i\rho_{i}$ suddenly falls down to -13.201 ($Q_{av}$ = -0.7334) where SkX emerges, then immediately increases to -7.568 ($Q_{av}$ = -0.420) at $T/{\cal J}$ = 3.1 where the SkX is considerably deformed as shown in Figure 2.

While $B_z$ = 0.28 T and   $T/{\cal J}$ = 2.4, eight skyrmions are observed. But  below this temperature, four  skyrmions and four bimerons appear. If one bimeron is considered to be composed of two semi-skyrmions, so the total  skyrmion number shown in Figure 6(b) is approximately  eight, which gives $Q_{av}$ = -0.830 for one skyrmion complex, that is close to those values tabulated in Table 1.

\emph{
More interestingly, the topological charge density calculated with Eq.(\ref{rhoQ}) for every SkX  also forms periodical and symmetric lattice which looks almost identical to the corresponding magnetic SkX, as displayed in Figure S1 of the supplementary part,  demonstrating the correctness of our simulations.}

Obviously, when the system is  polarized by a strong external magnetic field to become completely ferromagnetic, the averaged helicity per site  $\gamma$  is equal to zero  as shown in Figure 7(c). In the helical state, $\gamma$ is around -0.8313, and in the case of SkX texture,  the helicity is found to be in the region  -0.7822 $ < \gamma < $ -0.4072. The points  with $\gamma$ = 0 form the boundary between the ferromagnetic and chiral  phases.

\vspace{0.3cm}
\section{Conclusions and Discussion}
We have employed  a quantum simulation approach to investigate a quasi-2D Fe$_{0.5}$Co$_{0.5}$Si-like  ferromagnetic system. In the absence of external magnetic field, the helical texture  is the ground state.  When a moderate  external magnetic field is exerted perpendicular to the monolayer, SkX textures of the HCP structure are induced.   As the magnetic field strength is increased, the SkX textures  can be deformed, rotated, and the periodical lengths  increased.  Afterwards, complicated spin structures, such as skyrmions plus bimerons, finite-length helical strips,  sparse skyrmions surrounded by vortices, appear successively. And finally, the system becomes ferromagnetic  as $B_z \geq$ 0.86 T.

 In each SkX,  the spins of a  skyrmons  align clockwise in the $xy$-plane,  while those of the  neighboring shallow vortices  order  anti-clockwise in the same plane, so that the total (free) energy of the whole system can be  minimized and the SkX texture stabilized.  All chiral spin textures are right-handed since $D$ is positives.  Moreover, the  calculated topological charge density for every SkX  also forms periodical and symmetric lattice which is almost identical to the corresponding  SkX, demonstrating the correctness of our simulations.

In a zero or weak magnetic field, the periodic wavelengths  of the helical and SkX textures agree roughly with  the theoretical values estimated with Eq.(\ref{avelength}).  Late on, we found that \cite{LiuIan19}, when a weak a magnetic field is applied,  another formula
\begin{equation}
\tan\left(\frac{2\pi}{\lambda}\right)=\frac{D}{\cal J}\;
\label{wl2}
\end{equation}
 agrees better with the SkX periodicity in the diagonal  [110] or [-110] direction, as shown in Figure 2. However, this formula was actually derived for the helical state based on a continuous model \cite{Leonov}. As indicated by the authors, for the skyrmion states, no analytical expression for the periodicity  can be obtained \cite{Leonov}. Nevertheless, if the wavelength  estimated from Eq.(\ref{avelength}) is projected onto the diagonal direction, it gives  a value that is roughly equal to the periodicity evaluated from Eq.(\ref{wl2}).

As $B_z$ is enhanced, the periodicity is increased. Consequently, the lattice size chosen for simulation cannot fit the spin texture any longer, this mismatch can destroy the periodical structures in simulations even they exist in reality. Therefore, in order to generate symmetric and periodic spin textures and to study the physical properties accurately in strong external magnetic field, the lattice size has to be adjusted. In principle, a lattice which gives the lowest total (free) energy is expected to produce the best results.

An isolated skyrmion can be  induced from the helical state when the external magnetic field is stronger than the critical value $ B_{c1} = 0.2D^2/{\cal J}$ \cite{Kwon13, Kwon12,Han}. Inserting the values of ${\cal J}$ and  $D$ we use here into this formulae provides $B_{c1}$ = 0.148 T, which is comparable to   $B_{c1}$ = 0.11 T that we find in  our simulations. On the other hand, if $B$ exceeds another critical value $B_{c2} = 0.8D^2/{\cal J}$, SkX texture disappears,  the  skyrmion plus bimeron  states  emerge  afterwards \cite{Kwon13,Kwon12,Han}.  The above two formulas gives the ratio  $B_{c1}/B_{c2}$ = 0.25. However, for FeGe and Fe$_{0.5}$Co$_{0.5}$Si, the ratios of the two transition fields are observed  in experiments to be within a range of  (0.3, 0.39). In the present case, we find in simulations that $B_{c2}$ = 0.28 T, the ratio $B_{c1}/B_{c2} $ is approximately 0.39, which agrees well with the   experimental observation.

In chiral magnetic systems, the SkX wavelengths  are usually quite large. To calculate and display their periodical spin textures, the grid model has to be used,  where ${\cal J}$ and $D$ parameters  are scaled, so are other physical quantities
\cite{Kwon13,Kwon12}. To do this, we further quantize the grid with a quantum spin. From the calculated results presented above, we can conclude that this treatment is quite effective in describing the detailed spin textures  of the quasi-2D system, as they evolve, driven by changing  temperature and  external magnetic field.

Especially, the skymions  recently observed at interfaces are only a few nanometers in diameter \cite{Dupe2,Hagemeister16,Hagemeister,Kwon13}, and it is these magnetic materials that are of great importance in future technology. Obviously, within these magnetic systems, the spin vectors cannot change continuously, thus our discrete quantum model is expected to be more accurate and effective than those popularly used classical ones.

 \vspace{0.5cm}
 \centerline{\bf Acknowledgement}
{\small Z.-S. Liu acknowledges   the financial support provided by National Natural Science Foundation of China  under grant No.~11274177 and by University of Macau. H. Ian is supported by FDCT of Macau under grant 065/2016/A2 and by National Science Foundation of China under grant 11404415.}

\end{document}